\documentclass[10pt,conference]{IEEEtran}
\IEEEoverridecommandlockouts

\usepackage{tikz}
\usepackage{amsmath}
\usepackage{enumitem}
\usepackage{pifont}
\usepackage{multirow,makecell}
\usepackage{tcolorbox}
\usepackage{color,xcolor}
\usepackage{listings,amsfonts}
\usepackage{caption}
\usepackage{subcaption}
\usepackage{threeparttable}
\usepackage{bbding}
\usepackage{graphicx}
\usepackage{booktabs} 
\usepackage{longtable}
\usepackage{flushend}
\usepackage{wasysym}

\AtEndPreamble{
	\usepackage{hyperref}

	\hypersetup{
		colorlinks = true,
		linkcolor = purple,
		anchorcolor = purple,
		citecolor = purple,
		filecolor = purple,
		urlcolor = purple
	}
}

\newcommand{\tool}{\textsc{MiCoScan}}

\begin{document}

\title{Demystifying Cookie Sharing Risks in WebView-based Mobile App-in-app Ecosystems}

\author{
\IEEEauthorblockN{Miao Zhang\textsuperscript{\textasteriskcentered}\textsuperscript{\ding{41}}$^{1}$, Shenao Wang\textsuperscript{\textasteriskcentered \textdagger}$^{2}$, Guilin Zheng$^{1}$, Yanjie Zhao\textsuperscript{\textdagger}$^{2}$, Haoyu Wang\textsuperscript{\textdagger \ding{41}}$^{2}$
}
\IEEEauthorblockA{
$^{1}$ Beijing University of Posts and Telecommunications, China \\
$^{2}$ \textsuperscript{\textdagger}Huazhong University of Science and Technology, China\\
}
}

\maketitle

\begingroup\renewcommand\thefootnote{\textasteriskcentered}
\footnotetext{Both authors contributed equally to this research.}
\endgroup

\begingroup\renewcommand\thefootnote{\ding{41}}
\footnotetext{Miao Zhang~(zhangmiao@bupt.edu.cn) and Haoyu Wang~(haoyuwang\\@hust.edu.cn) are the corresponding authors.}
\endgroup

\begingroup\renewcommand\thefootnote{\textdagger}
\footnotetext{The full name of the affiliation is Hubei Key Laboratory of Distributed System Security, Hubei Engineering Research Center on Big Data Security, School of Cyber Science and Engineering, Huazhong University of Science and Technology.}
\endgroup

\begin{abstract}
Mini-programs, an emerging mobile application paradigm within super-apps, offer a seamless and installation-free experience. However, the adoption of the web-view component has disrupted their isolation mechanisms, exposing new attack surfaces and vulnerabilities. In this paper, we introduce a novel vulnerability called \textbf{Cross Mini-program Cookie Sharing~(CMCS)}, which arises from the shared web-view environment across mini-programs. This vulnerability allows unauthorized data exchange across mini-programs by enabling one mini-program to access cookies set by another within the same web-view context, violating isolation principles.
As a preliminary step, we analyzed the web-view mechanisms of four major platforms, including WeChat, AliPay, TikTok, and Baidu, and found that all of them are affected by CMCS vulnerabilities. These findings were responsibly disclosed and acknowledged with two CVEs. Furthermore, we demonstrate the collusion attack enabled by CMCS, where privileged mini-programs exfiltrate sensitive user data via cookies accessible to unprivileged mini-programs.
To measure the impact of collusion attacks enabled by CMCS vulnerabilities in the wild, we developed \tool{}, a static analysis tool that detects mini-programs affected by CMCS vulnerabilities. \tool{} employs \textit{web-view context modeling} to identify clusters of mini-programs sharing the same web-view domain and \textit{cross-webview data flow analysis} to detect sensitive data transmissions to/from web-views. Using \tool{}, we conducted a large-scale analysis of 351,483 mini-programs, identifying 45,448 clusters sharing web-view domains, 7,965 instances of privileged data transmission, and 9,877 mini-programs vulnerable to collusion attacks.
Our findings highlight the widespread prevalence and significant security risks posed by CMCS vulnerabilities, underscoring the urgent need for improved isolation mechanisms in mini-program ecosystems.
\end{abstract}

\section{Introduction}
A new mobile application paradigm, referred to as ``app-in-app''\cite{lu2020demystifying} or mini-programs\cite{yang2022cmrf,w3c,wang2023miniprogram}, has gained significant traction and popularity over the past few years. Under this paradigm, a host app or super-app runs a suite of sub-apps as in-app components through a JavaScript engine. This compact paradigm offers users a seamless and installation-free experience, eschewing the traditional process of downloading standalone mobile applications~\cite{wang2024miniscope,wang2024minichecker}. Within the super-app environment, users can access a diverse range of functionalities, such as shopping, dining, and transportation services, without leaving the host application. This integrated and cohesive experience enhances user engagement and stickiness. As a result, mini-programs have rapidly emerged as a popular and lightweight alternative to conventional mobile application development~\cite{han2024policy,deng2024counterfeit}. Currently, the most prominent and representative super-app, WeChat, boasts an ecosystem of over 3.5 million mini-programs, attracting over 600 million active users on a daily basis~\cite{zhang2022identity}. 

\begin{figure}[t]
    \centering
    \includegraphics[width=0.85\linewidth]{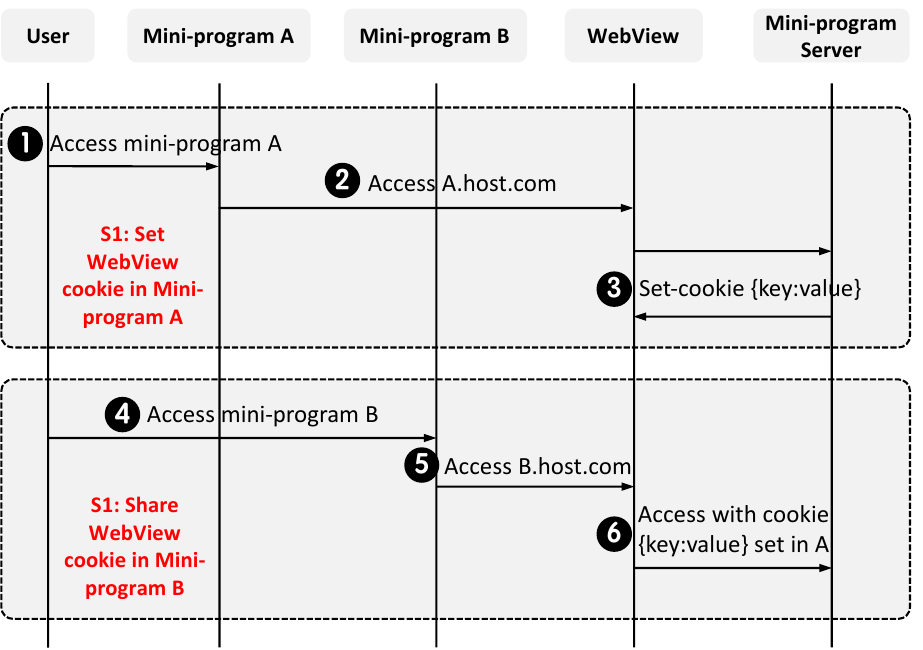}
    \caption{Cross Mini-program Cookie Sharing via Web-View.}
    \label{fig:cookie-sharing}
\end{figure}

As super-apps increasingly resemble operating systems~\cite{yang2023sok}, they are tasked with providing resource management~\cite{zhang2024minible,yang2023muid}, permission control~\cite{wang2024miniscope,wang2024rootfree}, and security mechanisms~\cite{wang2023browser,zhang2023trusted,han2023security} for the hosted mini-programs. Similar to traditional Android applications~\cite{android}, mini-programs employ isolation mechanisms~\cite{sandbox,yang2023sok,processiso} to ensure security and prevent unauthorized data sharing. This isolation model aims to create distinct sandboxes for each mini-program, restricting their access to system resources and user data based on predefined permissions and policies.
However, the introduction of the web-view component~\cite{webview} has disrupted the isolation model, introducing new attack surfaces and potential vulnerabilities. 
When mini-programs leverage the web-view component to render web content or access web resources, they inherit the cookies associated with that particular web-view instance. This inheritance mechanism, while intended for legitimate purposes such as maintaining user sessions, inadvertently opens a channel for unauthorized data exchange between mini-programs.

In this paper, we introduce a new type of vulnerability, called \textbf{Cross Mini-program Cookie Sharing~(CMCS)}. As illustrated in \autoref{fig:cookie-sharing}, the vulnerability arises from the shared web-view environment among mini-programs. When mini-program A accesses a web resource~(e.g., \texttt{A.host.com}) through the web-view component, it sets cookies within the current context~(\ding{182}-\ding{184}). Subsequently, when mini-program B accesses a different web resource (e.g., \texttt{B.host.com}) through the same web-view instance, it inherits and can access the cookies previously set by mini-program A~(\ding{185}-\ding{187}). This unintended sharing of cookies across mini-programs within the same web-view context violates the isolation principles and enables unauthorized data exchange.
As the preliminary analysis in our study, we investigated the web-view mechanisms of four major mini-program platforms, including WeChat, AliPay, TikTok, and Baidu, to assess their susceptibility to CMCS vulnerabilities. Our findings reveal that all four platforms are affected due to their shared web-view environments. We responsibly disclosed these vulnerabilities, and our efforts were acknowledged with 2 CVEs.

Specifically, based on the current implementations of web-view cookies, we demonstrate a type of attack enabled by CMCS, referred to as \textit{Collusion Attacks}. \textit{Collusion Attacks} involve a privileged mini-program obtaining sensitive user data by abusing its granted permissions. The privileged mini-program then transmits this sensitive data to the web-view and stores it in the web-view cookies. These cookies can subsequently be accessed by an unprivileged, colluding mini-program, effectively allowing the unauthorized sharing of privileged data that the latter should not have access to. 

To measure the potential impact of collusion attacks enabled CMCS vulnerabilities in the wild, we have developed \tool{}, a static analysis tool designed to detect vulnerable mini-programs. \tool{} employs a three-step approach:  
First, \tool{} performs \textit{web-view context modeling} by examining the URLs accessed through web-view components across different mini-programs. Mini-programs that interact with the same web-view domain are grouped into clusters, as these groupings represent contexts where cookies may be shared, thereby exposing them to potential vulnerabilities.  
Next, \tool{} introduces \textit{cross-webview data flow analysis} to inspect the bi-directional communication channels between mini-programs and their associated web-view components. It detects instances where privileged data, such as user contacts, location, or other sensitive information, is transmitted from mini-programs to the web-view environment. Such transmissions increase the likelihood of privileged data being stored in web-view cookies, where it becomes accessible to other mini-programs within the same context.  
Finally, by integrating web-view context modeling with privileged data transmission inspection, \tool{} effectively identifies mini-programs susceptible to CMCS vulnerabilities.

Based on \tool{}, we conducted a large-scale, systematic analysis of 351,483 real-world mini-programs, offering the first comprehensive measurement of CMCS vulnerabilities in the wild. Our investigation identified 45,448 unique clusters of mini-programs accessing the same web-view domain. Further examination revealed 7,849 instances of mini-programs transmitting privileged data (such as contacts, location, and authentication tokens) to web-views and 116 instances of mini-programs receiving privileged data from web-views, indicating bi-directional data flows that compromise isolation. Through systematic analysis of the associated cookies, we identified 9,877 mini-programs vulnerable to collusion attacks, highlighting the widespread impact of this issue.

\noindent \textbf{Contributions.} The contributions are outlined as follows:

\begin{itemize}[leftmargin=15pt]
    \item \textbf{Novel Attack Vectors~(\autoref{sec:attacks})}. We are the first to systematically study the isolation mechanisms in mini-programs. Our research uncovered the CMCS vulnerability in four major platforms, including WeChat, AliPay, TikTok, and Baidu, which enables unauthorized data sharing across mini-programs. These findings have been responsibly disclosed and acknowledged with 2 CVEs.
    
    \item \textbf{Practical Detection~(\autoref{sec:detection})}. We present \tool{}, a static analysis tool that detects potential CMCS vulnerabilities through. The web-view context modeling identifies clusters sharing the same web-view domain as potential candidates. The cross-webview data flow analysis examines bi-directional communication channels to uncover sensitive data transmission.

    \item \textbf{Large-scale Measurement~(\autoref{sec:evaluation})}. We conduct a large-scale study of 351,483 mini-programs, revealing the widespread prevalence of the CMCS vulnerability. Our analysis identifies 45,448 clusters of mini-programs sharing web-view domains and 7,965 instances of privileged data transmission to/from web-views, leading to the discovery of 9,877 vulnerable mini-programs.
\end{itemize}

\section{Background}
\subsection{Mini-programs}
\label{sec:architecture}

\begin{figure}[t]
    \centering
    \includegraphics[width=0.9\linewidth]{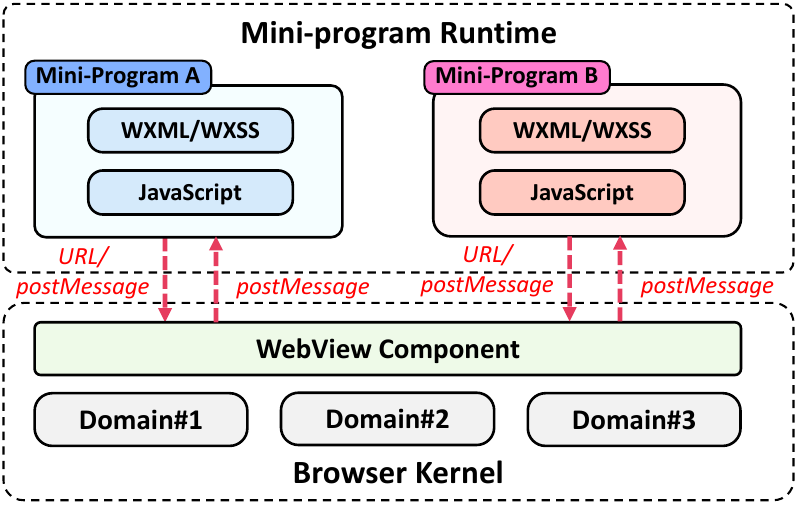}
    \caption{Architecture of Mini-programs.}
    \label{fig:architecture}
\end{figure}

\noindent \textbf{Architecture of Mini-programs.}
A mini-program typically comprises two integral components: the front end, executing within the host program, and the back end, operating on the server to provide essential services. Similar to how Android necessitates an APK for application usage, the host program must acquire the \texttt{WXAPKG} package before launching the mini-program. As illustrated in \autoref{fig:architecture}, this package contains crucial resources for the mini-program, including \texttt{WXML} files, \texttt{WXSS} files, and \texttt{JavaScript} files. \texttt{WXML} files, akin to \texttt{HTML}, define the UI interface structure, while \texttt{WXSS} files, resembling \texttt{CSS}, dictate the UI's style. \texttt{JavaScript} files are responsible for implementing the interactive logic, managing user interactions, handling network requests, and more. Additionally, the package may include various resource files, such as images.

\noindent \textbf{Web-View Components.}
The web-view component is provided by WeChat for developers to introduce HTML5~(H5 for short) pages. It contains four attributes: \texttt{src}, \texttt{bindmessage}, \texttt{bindload}, and \texttt{binderror}. The \texttt{src} attribute points to the embedded URL of web-view pages, which needs to be configured in the business domain name of the mini-program's backend. The remaining three attributes serve as event handlers: The \texttt{bindmessage} attribute is used to bind an event handler for receiving messages sent from the embedded web-view page to the mini-program, while \texttt{bindload} and \texttt{binderror} are callback functions triggered when the web page loads successfully or fails to load.

\noindent \textbf{Bi-direction Communication.}
One of the key features of the web-view component is its ability to facilitate bi-directional communication between the mini-program and the embedded web-view page, as illustrated in \autoref{fig:architecture}. This communication is primarily managed through the \texttt{postMessage} API and the \texttt{bindmessage} attribute. 

\begin{itemize}[leftmargin=15pt]
    \item \textbf{C\#1: Mini-program$\to$Web-View.} Two prevalent methods are transmitting information through URL parameters and using the \texttt{postMessage} API: 1) One straightforward approach involves attaching the required data as query parameters to the URL that the Web-View component loads. When the web-view loads the URL, JavaScript within the web page can extract these parameters. 2) A more sophisticated and flexible approach is to use the \texttt{postMessage} API. This involves creating a web-view context using \texttt{createWebViewContext} and then invoking the \texttt{postMessage} method to send messages.  The web-view page can listen for these messages using the \texttt{onMessage} event listener.

    \item \textbf{C\#2: Web-View$\to$Mini-program.} The \texttt{postMessage} also allows the web-view page to send messages to the mini-program. On the mini-program side, the \texttt{bindmessage} attribute is used to define an event handler that listens for these messages. When the web-view page sends a message, the event handler in the mini-program is triggered, allowing the mini-program to process the message content and respond accordingly.
\end{itemize}

\subsection{Security Model and Isolation}
The security model of mini-programs is a fundamental aspect that ensures the integrity, privacy, and stability of the host application and its contained mini-programs. This model is built upon a robust isolation mechanism that segregates each mini-program into an independent operational environment. This segregation prevents unauthorized access and interference between mini-programs and protects the host application from potential security breaches. The specific implementation of this technology includes:

\begin{itemize}[leftmargin=15pt]
    \item \textbf{Sandbox Technology.} Sandbox technology~\cite{sandbox} confines the operations of a mini-program within a controllable security sandbox. This restriction limits the mini-program's interface calling permissions, allowing access only to interfaces provided by the host application. This containment ensures that mini-programs can only interact with permitted system components, enhancing security.
    
    \item \textbf{Independent Processes.} Each mini-program in the host application operates as a separate process~\cite{processiso}. During execution, each mini-program functions as an individual process, ensuring that any abnormalities or crashes within one mini-program do not impact the operation of others. This process isolation enhances the stability and reliability of the entire system.
    
    \item \textbf{Resource Isolation.} The host application allocates an independent resource directory for each mini-program~\cite{sandbox}. Resources saved by each mini-program are stored in isolated directories, ensuring they are not shared with or accessible to other mini-programs. This resource isolation maintains the independence of each mini-program's data.
\end{itemize}

\subsection{Cookie Mechanism}
In the context of mini-programs and web-view components, the cookie mechanism is crucial for ensuring seamless user experiences and consistent data management.

\noindent \textbf{Cookies in Mini-Programs.}
Unlike traditional web applications, mini-programs operate within a controlled environment managed by the host application. The cookie mechanism in mini-programs is tailored to this unique context, emphasizing isolation and security. Each mini-program operates within its own sandboxed environment.  This means that cookies set by one mini-program are stored in a separate storage space that is inaccessible to other mini-programs. In addition to isolating cookies between mini-programs, there is also a clear separation between cookies used by mini-programs and those used by the host application.  This ensures that cookies set by mini-programs do not interfere with the host's data and vice versa.

\noindent \textbf{Cookies in Web-View Components.}
Web-view components serve as embedded browser engines within mini-programs, enabling the display of web content. The cookie mechanism in web-view components mirrors that of traditional web browsers. Contrary to the isolation seen in many aspects of mini-program architecture, the cookies managed within web-view components and the host application are not isolated. Instead, they share the same embedded browser instance, leading to shared cookie storage~(which we will further discuss in \autoref{sec:root-cause}). This shared environment introduces a series of privacy and security risks that must be carefully managed\footnote{Our primary focus in this study will be on the cookies used within web-view components unless otherwise specified.}. 
\section{The CMCS Attacks}
\label{sec:attacks}

While the host application offers various methods to isolate mini-programs, the introduction of the web-view component disrupts the original security model. Notably, we observed that when external web pages within the web-view components of two distinct mini-programs belong to the same domain, these two web pages, belonging to different mini-programs, can share cookies during runtime. Now we present the CMCS attacks, whereby an attacker can exploit this vulnerability to facilitate the collusion attacks.

\subsection{Threat Model and Scope}

% 如图 1 所示，跨小程序cookie共享涉及多方，例如主机应用程序（例如微信）、以及主动的cookie设置方和被动的cookie共享方小程序的前端及其后端。在研究 cookie 共享攻击时，我们基于几个关键假设。我们假设主机应用程序允许多个小程序嵌入 web-view 组件，这些组件可以加载和呈现外部网页。此外，我们假设这些 web-view 组件在 cookie 存储方面不会严格隔离不同的小程序，特别是当网页属于同一域时。以上假设我们会在后文Section中进行证明。

\noindent \textbf{Assumption.}
As illustrated in \autoref{fig:cookie-sharing}, cross mini-program cookie sharing involves multiple parties, including the host application (e.g., WeChat), and both the cookie-setting mini-program and the cookie-receiving mini-program, as well as their respective front-ends and back-ends. Our study of CMCS attacks is based on several key assumptions. First, we assume that the host application permits multiple mini-programs to embed web-view components, which can load and render external webpages. This setup is critical for enabling the interaction between different mini-programs via shared web-view components. Second, we assume that these web-view components do not enforce strict isolation regarding cookie storage between different mini-programs, particularly when the webpages belong to the same domain\footnote{It is worth noting that the Same-Origin Policy (SOP) is not a strong security assumption in this context, as many mini-programs are developed based on frameworks or hosted by third-parties, making the presence of third-party domains within mini-programs highly common.}. This lack of isolation allows cookies to be shared across mini-programs, creating a potential vector for attacks. The validity of these assumptions will be substantiated in \autoref{sec:root-cause}.

% 鉴于这些假设，攻击者主要在小程序生态系统范围内拥有特定能力。攻击者可以在主机应用程序的生态系统内开发和分发恶意小程序。此恶意小程序旨在从攻击者控制的域加载网页。攻击者的主要能力在于他们控制托管在其域上的网页的内容和行为。此功能包括在恶意小程序的 Web 视图组件中读取、写入和操纵 Cookie 的能力。因此，攻击者可以利用共享的 Cookie 来跟踪不同小程序之间的用户活动，从而侵犯用户隐私并实现跨小程序跟踪。我们的威胁模型的范围仅限于单个主机应用程序中的小程序和 Web 视图组件之间的交互。我们专注于 Cookie 共享的安全隐患，特别是它如何导致隐私泄露和未经授权的用户跟踪。该威胁模型不包括 Web 视图组件交互范围之外的其他形式的进程间通信或潜在漏洞。

\noindent \textbf{The Attacker's Capability.}
Given these assumptions, the attacker possesses specific capabilities primarily within the scope of the mini-program ecosystem. The attacker can develop and distribute a malicious mini-program within the host application's ecosystem. This malicious mini-program is designed to load a webpage from a domain under the attacker's control. The attacker's main capability lies in their control over the content and behavior of the webpage hosted on their domain. This capability includes the ability to read, write, and manipulate cookies within the web-view component of the malicious mini-program. Consequently, the attacker can leverage shared cookies to track user activity across different mini-programs, thereby breaching user privacy and enabling cross-mini-program tracking. The scope of our threat model is confined to the interaction between mini-programs and web-view components within a single host application. We focus on the security implications of cookie sharing, particularly how it can lead to privacy breaches and unauthorized tracking of users. This threat model does not encompass other forms of inter-process communication or potential vulnerabilities outside the realm of web-view component interactions.

% 小程序范例已得到多个超级应用程序的支持。为了说明Cookie Sharing在不同App-in-app生态中的普遍性，我们对微信，支付宝，百度，抖音，快手等主要的超级应用平台进行了概念验证。而在Detection和Measurement Study中，我们重点关注微信中的小程序，因为微信小程序范式的先去，已经托管了超过340万款小程序。与其他平台相比，针对微信的测量性研究更能突出Cookie Sharing在野的利用现状以及造成的影响。但是请注意，尽管我们在本文中仅对微信进行了测量研究，但重要的是要强调由于小程序生态的标准化架构，我们提出的攻击和检测方法到其他平台的可扩展性和可转移性。

\noindent \textbf{Scope.} 
The app-in-app ecosystem has gained significant attention in recent years. To illustrate the existence and applicability of cookie sharing across different super-app platforms, we conducted a preliminary proof-of-concept study on four major platforms: WeChat, Alipay, Baidu, and TikTok. This initial analysis focused on evaluating the web-view mechanisms of these platforms to confirm their susceptibility to CMCS vulnerabilities. Building on these findings, our subsequent measurement study specifically targets WeChat. As a pioneer in the mini-program paradigm, WeChat hosts over 3.5 million mini-programs~\cite{zhang2022identity}, making it an ideal candidate for large-scale analysis to evaluate the real-world impact of CMCS vulnerabilities. It is important to note that while our measurement is confined to WeChat, the standardized architecture~\cite{w3c} of mini-program ecosystems ensures that our proposed methods are scalable and transferable to other platforms.

\subsection{The Design Pitfalls and Root Cause}
\label{sec:root-cause}

In this section, we report our preliminary analysis of cookie management within app-in-app ecosystems. Our study reveals that all four major ecosystems, including WeChat, Alipay, TikTok, and Baidu, exhibit CMCS vulnerabilities. 

\begin{figure}[t]
    \centering
    \includegraphics[width=0.85\linewidth]{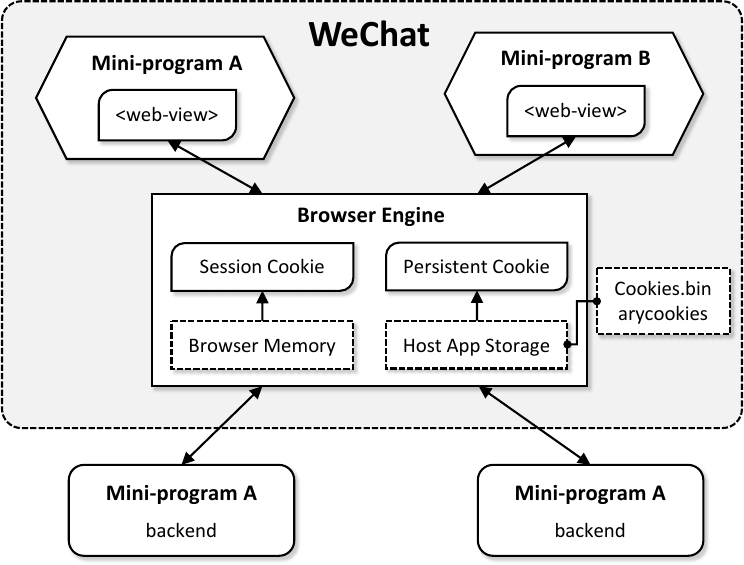}
    \caption{Design Pitfalls and Root Cause of CMCS.}
    \label{fig:cookie-persistence}
\end{figure}

\noindent \textbf{Design Pitfalls of WeChat.} 
We used Safari and the iOS version of WeChat for remote debugging of web pages within web-view components. Our observations revealed two types of cookies: (1) persistent cookies (with explicit expiration times denoted by the \texttt{Expires} field); and (2) session cookies (with the \texttt{Expires} field set to the session, indicating they last for the session duration). Through analysis of captured network packets, we discovered that both types of cookies are shared across different mini-programs' web-view components, suggesting that these web-view components share the same browser kernel. After exiting a web-view component and closing the mini-program, session connections do not immediately disconnect. Consequently, when opening another mini-program with an external web page, session cookies are shared. 
As illustrated in \autoref{fig:cookie-persistence}, we identified the \texttt{Cookies.binarycookies} file located under \texttt{`/Library/Cookies/'} in the WeChat directory of iOS, which contains cookies generated by the web-view components of different mini-programs and is stored in a shared public directory rather than isolated storage. This confirms that cookies are shared across the web-view components of different mini-programs, highlighting a violation of the intended isolation principles.
Similarly, on Android, we observed the same behavior with web-view components. The cookies associated with external web-view are stored at \texttt{app\_xweb\_data/xweb\_xxx/profile/Cookies} in the WeChat directory. This further confirms that web-view components across different mini-programs on both iOS and Android share the same underlying cookie storage mechanism.

\begin{table}[t]
\fontsize{8.5}{11}\selectfont
\centering
\caption{CMCS in Four Major Mini-program Ecosystems.}
\label{table:platforms}
\begin{tabular}{c||c||c}
\hline
\textbf{Ecosystem} & \textbf{Cookie Storage}    & \textbf{CVE} \\ 
\hline
WeChat~\cite{wechat}    & Cookies.binarycookies &    CVE-2024-40433   \\
AliPay~\cite{alipay}    & Cookies.binarycookies &    pending   \\
TikTok~\cite{tiktok}    & Cookies.binarycookies &    CVE-2025-25436  \\
Baidu~\cite{baidu}     & Cookies.binarycookies &    pending   \\ 
\hline
\end{tabular}
\end{table}

\noindent \textbf{CMCS in Other Mini-program Ecosystems.} As illustrated in \autoref{table:platforms}, we also analyzed the iOS and Android versions of Alipay~\cite{alipay}, TikTok~\cite{tiktok}, and Baidu~\cite{baidu}. We found that, like WeChat~\cite{wechat}, the cookies generated by the external web pages of the web-view components of their mini programs are stored in the corresponding \texttt{/Library/Cookies/Cookies.binarycookies} path in the application directory, and during the access to the mini-program, we found the same situation that appeared in WeChat, that is, some cookies of the domain name in the web-view will be saved In the \texttt{Cookies.binarycookies} file, that is to say, the same cookie sharing will also occur in these mini-program ecosystem. Based on these findings, we responsibly disclosed the CMCS vulnerabilities to the affected platforms. As shown in \autoref{table:platforms}, our disclosures have been acknowledged with two CVEs. 

\subsection{Attack Scenario}

\begin{figure}[t]
    \centering
    \includegraphics[width=0.9\linewidth]{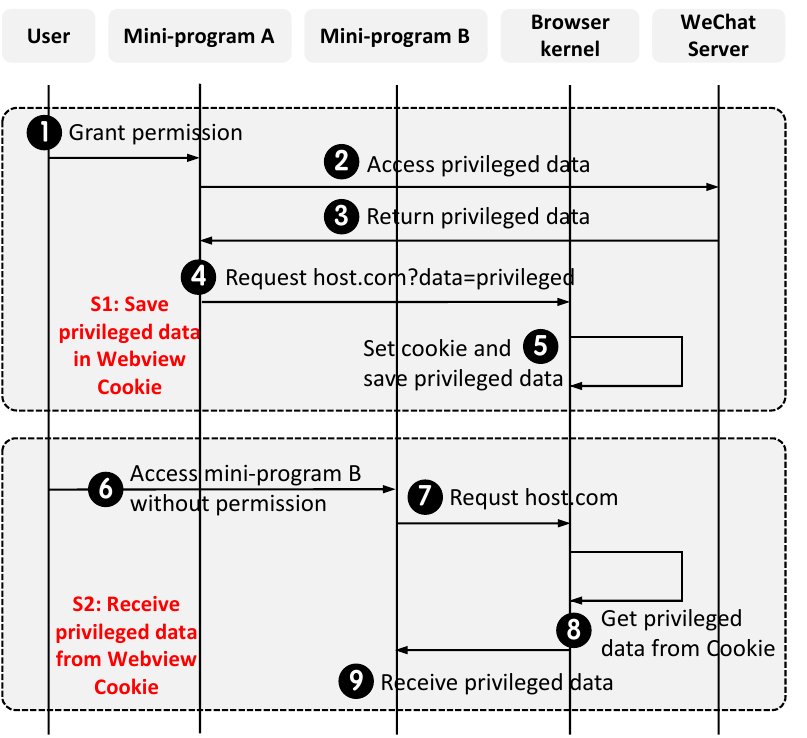}
    \caption{Collusion Attack via CMCS.}
    \label{fig:collusion-attack}
\end{figure}

\noindent \textbf{Collusion Attack.} 
One potential attack scenario is the collusion between mini-programs that leverage cookie sharing to share privileged data. In such scenario, one mini-program A can maliciously exploit shared cookies to access privileged data initially intended for another mini-program B, thereby circumventing established permission controls. We explain the mechanism of such an attack in \autoref{fig:collusion-attack}, using a hypothetical example involving two mini-programs, A and B, within a platform like WeChat.
Firstly, consider that a user grants mini-program A the necessary permissions to access privileged data~(\ding{182}\ding{183}). Mini-program A retrieves this data~(\ding{184}) and saves it within the web-view's cookie~(\ding{185}\ding{186}). This storage action is performed without the user's explicit awareness that such sensitive data is being saved in a shared, less secure manner. Subsequently, the user accesses mini-program B, which does not have the same permissions as mini-program A to access the privileged data directly. However, because mini-program B operates within the same web-view environment, it can exploit the shared cookie mechanism. When mini-program B requests the same web-view, it can read the cookies stored by mini-program A. This allows mini-program B to retrieve the privileged data without the user's explicit consent or knowledge. This process effectively bypasses the intended security model, which relies on explicit permissions. The shared web-view component acts as an unintended bridge, enabling data leakage from a privileged context to an unprivileged one. 

\begin{figure*}[t]
    \centering
    \includegraphics[width=0.85\linewidth]{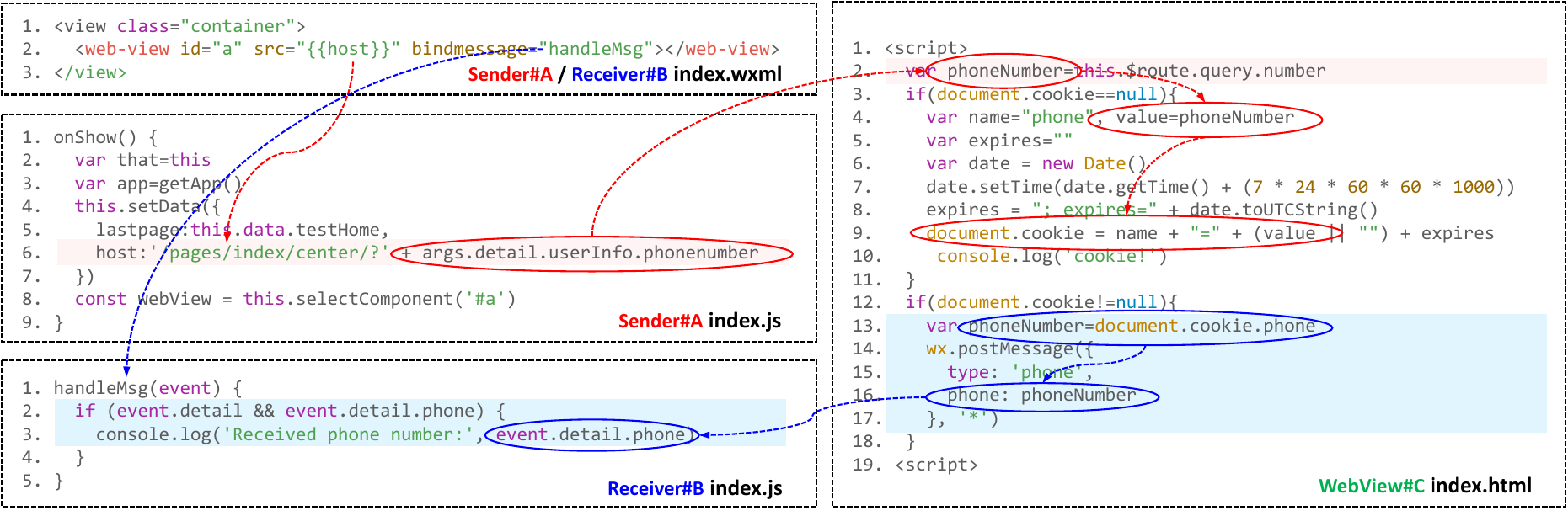}
    \caption{A Running Example Code Snippet of Collusion Attack via CMCS.}
    \label{fig:code-snippet}
\end{figure*}

\subsection{A Running Example}
The provided code snippets in \autoref{fig:code-snippet} demonstrate a collusive attack that enables cross mini-program sharing of sensitive user data through cookie sharing within the web-view component.
The attack involves two mini-programs, Sender\#A and Receiver\#B, and a shared web-view component Web-View\#C. The goal is to transmit the user's phone number from Sender\#A to Receiver\#B, even though Receiver\#B is not authorized to access this sensitive information directly.

\noindent \textbf{Sender\#A.}
In \texttt{index.wxml}, Sender\#A embeds a web-view component and binds its \texttt{src} attribute to the \texttt{host} variable, which is set to a URL containing the user's phone number \texttt{args.detail.userInfo.phonenumber}.

\noindent \textbf{Web-View\#C}
In \texttt{index.html}, upon receiving the URL containing the user's phone number, Web-View\#C extracts the phone number from the URL query parameter (\texttt{this.\$route.query.number}).
If no cookie is present, Web-View\#C creates a new cookie with the name \texttt{phone} and the value set to the extracted phone number \texttt{value}.
If a cookie already exists, Web-View\#C retrieves the phone number from the existing cookie (\texttt{document.cookie.phone}).
Web-View\#C then sends a \texttt{postMessage} back to the mini-program, containing the phone number.

\noindent \textbf{Receiver\#B}
In \texttt{index.wxml}, Receiver\#B also embeds a web-view component and binds its \texttt{bindmessage} attribute to the \texttt{handleMsg} function.
In \texttt{index.js}, the \texttt{handleMsg} function listens for \texttt{event} received from the web-view component.
If the received \texttt{event} contains a phone number (\texttt{event.detail.phone}), Receiver\#B logs it, thereby accessing sensitive user data from Sender\#A via CMCS.
\section{Design of \tool{}}
\label{sec:detection}
CMCS occurs especially when mini-programs operate under the same domain. This risk is highlighted by the potential for privileged data to be transmitted to the web-view and subsequently stored in cookies. So we need to identify the web-view domain context and privileged data flow. In this section, we introduce \tool{}, which identifies web-view contexts and employs cross web-view data flow analysis to pinpoint the transmission of privileged data during bidirectional communication.

\subsection{Step I: Web-View Context Modeling}
\label{sec:webview}
\begin{figure}[t]
    \centering
    \includegraphics[width=0.8\linewidth]{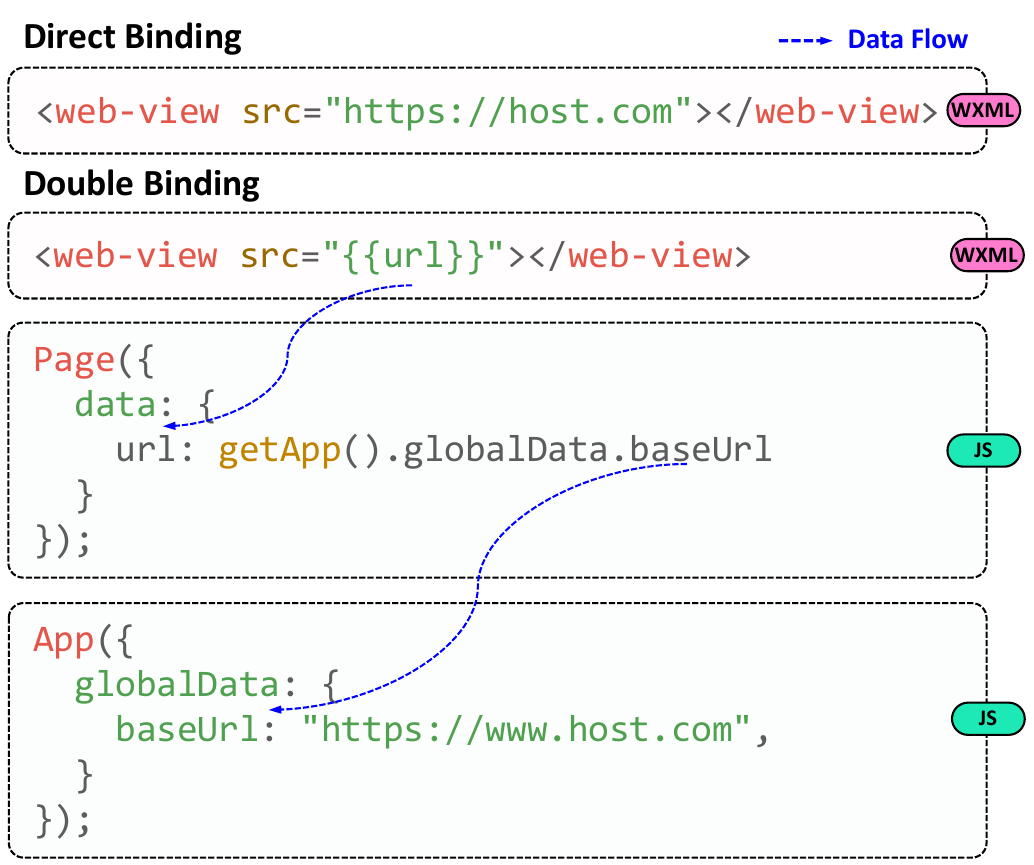}
    \caption{Data Binding Mechanism in Web-View Domain.}
    \label{fig:binding}
\end{figure}

The first step in detecting CMCS vulnerabilities is to accurately identify the domain context of each web-view instance within the mini-programs. This process involves resolving the two-way binding syntax used in mini-programs to determine the web-view \texttt{URL}. As illustrated in \autoref{fig:binding}, there are two primary methods for binding \texttt{URL} to the \texttt{src} attribute: \textit{direct binding} and \textit{double binding}. 
\begin{itemize}[leftmargin=15pt]
    \item \textbf{Direct Binding} involves specifying the \texttt{URL} directly within the \texttt{src} attribute of the web-view component. This method is straightforward and does not involve any dynamic data binding or JavaScript logic. When the user triggers the web-view component, the web-view component will jump directly to the URL pointed to. For the direct data binding, we can easily extract the target \texttt{URL} of the web page from the web-view component.
    
    \item \textbf{Double Binding}, also known as two-way binding, involves dynamically binding the \texttt{src} attribute to a JavaScript variable. This method allows the \texttt{URL} to be determined at runtime based on the application state or external data sources. Although the URL declaration is in the web-view component, the definition and assignment of data are in the JavaScript code. As shown in \autoref{fig:binding}, although the variable \texttt{url} in the web-view component is bound in \texttt{page.wxml}, its actual declaration is in \texttt{page.js}, and the data is assigned through \texttt{baseUrl} of \texttt{globalData}. So we can get the domain name of the \texttt{URL} based on data flow analysis of JavaScript and WXML files.
\end{itemize}

To resolve the target \texttt{URL} of the web-view component, the first step is to identify the bound variables in WXML, and then the next step is to trace their origins and transformations within the JavaScript code. This involves following the variable from its declaration to its final assignment, capturing all intermediate states and modifications. Although the data flow and value analysis follow the general flow of JavaScript program analysis, special handling is required for global variables in \texttt{Page} instances and \texttt{App} instances due to their unique assignment and referencing syntax supported by the host app. Traditional program analysis frameworks often struggle to support these nuances. In mini-programs, variables bound in WXML are typically located in the \texttt{data} object, which serves as a state storage akin to global variables within the current \texttt{Page} instance. The data object's properties can be dynamically updated via \texttt{this.setData} method, and referenced using \texttt{this.data.prop}. Additionally, a global \texttt{App} instance exists, encompassing the \texttt{globalData} object, which can be accessed and modified across any page using \texttt{getApp().globalData.prop}. Consequently, during data flow and value analysis, we need to parse these referencing and assignment rules to maintain a global variable table for the page of the mini-program, thereby accurately obtaining the values of variables bound in WXML.

\subsection{Step II: Cross Web-View Data Flow Analysis}

Another critical aspect of CSCS is the potential transmission of privileged data. To address this, we propose analyzing the data transmission during the bi-directional communication between the mini-program and the web-view. As described in \autoref{sec:architecture}, there are two channels for communication from the mini-program to the web-view: URL parameter concatenation and \texttt{postMessage} invocations. The communication from the web-view to the mini-program can only occur through \texttt{postMessage} calls, and the mini-program receives via the \texttt{bindMessage} handler.
Since the web-view logic executes on the browser engine, we can only examine the reception of privileged data from the web-view via the \texttt{bindMessage} handler. 
The data transmission using URL parameter concatenation can be analyzed based on Step I, so we focus on examining privileged data sent from \texttt{postMessage} invocations and received from \texttt{bindMessage} handler in this step.

\noindent \textbf{C\#1: Mini-program$\to$Web-View.} 
To scrutinize the potential privileged data transmission through \texttt{postMessage} calls, we employ a two-pronged approach: identifying \texttt{postMessage} invocations and tracing the data flow of messages.
The first step involves locating all instances of \texttt{postMessage} calls, which is achieved by traversing the Abstract Syntax Tree~(AST) and identifying function calls with the name \texttt{postMessage}. Additionally, we locate invocations of \texttt{createWebViewContext}, as this function is a prerequisite for \texttt{postMessage} communication with the web-view.
Once the \texttt{postMessage} call sites are identified, we trace the data flow of the messages being sent. This involves backward data flow analysis to determine the origins and transformations of the message content. Particular attention is paid to data sources that may contain privileged information, such as user credentials, device identifiers, or sensitive application data.

\noindent \textbf{C\#2: Web-View$\to$Mini-program.}
Complementing the analysis of \texttt{postMessage} data transmission, we also examine the reception of data from the web-view through \texttt{bindMessage} handlers within the miniprogram.
The analysis begins by identifying the callback functions bound to \texttt{bindMessage} within the WXML files. These callbacks serve as event handlers for messages received from the web-view.
Subsequently, we leverage AST parsing to locate the corresponding callback functions within the JavaScript code. Inside these functions, we identify instances where the received data is accessed through the \texttt{event.detail} property, which contains the message content sent from the web-view.
Once the access points of \texttt{event.detail} are identified, we trace the data flow of the received messages within the mini-program code. This involves forward data flow analysis to determine how the received data is processed, stored, and propagated throughout the application.

\subsection{Step III: Collusion Attack Detection}

To detect potential collusion attacks enabled by CMCS vulnerabilities, we combine the results from Step I and Step II to identify unauthorized data exchange between mini-programs.

\noindent \textbf{Grouping of Candidates.}  
The first step is to group mini-programs that share the same web-view domain. Using the domain context extracted in Step I, where the target URLs of web-view instances were resolved through static data flow analysis, we identify mini-programs operating within the same web-view domain. Mini-programs sharing the same domain inherently share cookies and other browser states, making them potential candidates for cross-mini-program cookie sharing vulnerabilities. By grouping these mini-programs together, we establish the scope of potential collusion and limit further analysis to interactions within each group. 

\noindent \textbf{Analysis of Overlapping Sensitive Data Flows.}  
Once mini-programs are grouped by web-view domain, the next step is to analyze sensitive data flows within each group to detect potential overlaps between the data sent and received by mini-programs. For every pair of mini-programs, we examine the data sent to the web-view by one mini-program and determine whether it can be accessed or processed by another mini-program in the same group. Specifically, we analyze the data sent to the web-view through \texttt{postMessage} invocations or URL parameter concatenations using backward data flow analysis. This helps trace the origins of transmitted data to identify whether it includes sensitive information, such as user credentials, device identifiers, or authentication tokens.
Similarly, we analyze the data received from the web-view by each mini-program through \texttt{bindMessage} handlers using forward data flow analysis. By correlating the data sent by one mini-program with the data received by another, we identify potential paths where sensitive information is exchanged between mini-programs. This pairwise analysis ensures that all possible interactions between mini-programs within the same domain group are systematically explored, enabling the detection of unauthorized data flows that could lead to collusion attacks.
\begin{figure*}[t]
    \centering
    \includegraphics[width=0.9\linewidth]{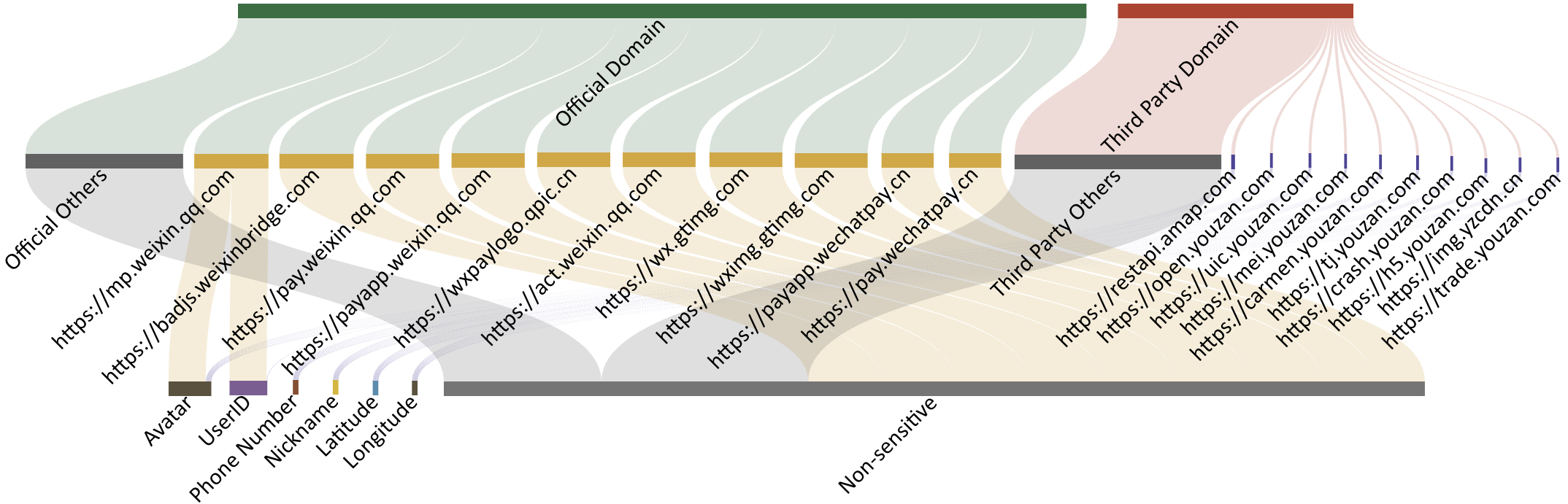}
    \caption{Privileged Data Exposure Across Web-View Domains in Mini-Programs.}
    \label{fig:overlap}
\end{figure*}

\section{Implementation}
\noindent \textbf{PoC Implementation.}
We have implemented Proof-of-Concept (PoC) code following the developer guidelines of super apps. As cookie sharing typically occurs within the same web-view domain, we developed a sender mini-program and a receiver mini-program that can access the same web-view domain.
For the collusion attack, we injected the privileged phone number obtained by the sender into a cookie. The non-privileged receiver then successfully retrieved this sensitive data from the cookie, bypassing the app's isolation mechanisms and enabling unauthorized data exposure.

\noindent \textbf{\tool{} Implementation.}
We have implemented \tool{} based on the existing mini-program data flow analysis framework \textsc{TaintMini}~\cite{wang2023taintmini}. Specifically, we added custom rules to track web-view domains accessed by each mini-program, as cookie sharing occurs within the same domain. We monitored URLs passed to web-view components and mapped mini-programs to their associated domains.
Additionally, we incorporated rules to detect potential transmission of privileged data like user identifiers between the mini-program and web-view. This analysis focused on data flows from sensitive sources (e.g., user info APIs) propagating to the web-view through communication channels.
By combining identified web-view domains and tracked privileged data flows, \tool{} can pinpoint instances where sensitive data may be shared across mini-programs via web-view cookies. These potential vulnerabilities are reported for further investigation and remediation.

\section{Measurement Results}
\label{sec:evaluation}
\subsection{Experimental Setup}
\noindent \textbf{Dataset.}
To collect mini-programs, we leveraged the open-source MiniCrawler~\cite{zhang2021measurement} to crawl and download package files from the WeChat mini-program ecosystem. Our initial dataset encompasses 415,550 mini-programs. However, after unpacking, only 351,483 mini-programs can be properly extracted and utilized for subsequent analysis.

\noindent \textbf{Running Environment.}
Our experiments are conducted on a server running Ubuntu Linux of 22.04 version with two 64-core AMD EPYC 7713 and 256 GB RAM. The static analysis leverages
the server’s computational capacity by utilizing 32 threads, enabling high parallelism for efficient processing of the large mini-program dataset.

\subsection{Vulnerability Prevalence}

\begin{table}[t]
\centering
\fontsize{9}{12}\selectfont
\begin{tabular}{lr}
    \hline
    \textbf{Filtering Methods} & \textbf{Count}\\
    \hline
    Total number of mini-programs & 351,483\\
    Filter\#1: Containing \texttt{<web-view>} & 261,538 \\
    Filter\#2: Web-View domain clustering & 45,448 \\
    Filter\#3: Clusters of mini-programs more than 5 & 3,109 \\
    % Filter 5: Manual analysis & 100 \\
    \hline
\end{tabular}
\caption{Filtering Process for Web-View Domain Analysis in Mini-Programs.}
\label{tab:classes}
\end{table}

\begin{table*}[t]
\centering
\fontsize{8}{10}\selectfont
\caption{Privileged Data Transmitted between Mini-program and Web-View. \textbf{S} Denotes Privileged Data Sent from Mini-program to Web-View, \textbf{R} Denotes Data Received from Web-View.}
\label{tab:privileged-data}
\begin{tabular}{c||>{\centering\arraybackslash}p{0.72cm}|>{\centering\arraybackslash}p{0.72cm}||>{\centering\arraybackslash}p{0.72cm}|>{\centering\arraybackslash}p{0.72cm}||>{\centering\arraybackslash}p{0.72cm}|>{\centering\arraybackslash}p{0.72cm}||>{\centering\arraybackslash}p{0.72cm}|>{\centering\arraybackslash}p{0.72cm}||>{\centering\arraybackslash}p{0.72cm}|>{\centering\arraybackslash}p{0.72cm}||>{\centering\arraybackslash}p{0.72cm}|>{\centering\arraybackslash}p{0.72cm}}
\hline
\multirow{2}{*}{\textbf{Categories}} & \multicolumn{2}{c||}{\textbf{Phone Number}} & \multicolumn{2}{c||}{\textbf{Nickname}} & \multicolumn{2}{c||}{\textbf{Latitude}} & \multicolumn{2}{c||}{\textbf{Longitude}} & \multicolumn{2}{c||}{\textbf{Avatar}} & \multicolumn{2}{c}{\textbf{User ID}} \\
\cline{2-13}
                                     & \textbf{S}       & \textbf{R}      & \textbf{S}        & \textbf{R}        & \textbf{S}        & \textbf{R}        & \textbf{S}         & \textbf{R}        & \textbf{S}       & \textbf{R}       & \textbf{S}       & \textbf{R}       \\ \hline
\textbf{Individual Seller}           & 772              & 14              & 353               & 2                 & 548               & /                 & 548                & /                 & 1,783            & 2                & 491              & 33               \\
\textbf{Life Service}                & 112              & /               & 72                & /                 & 175               & /                 & 175                & 3                 & 459              & /                & 133              & /                \\
\textbf{Tools}                       & 106              & /               & 33                & /                 & 6                 & /                 & 6                  & /                 & 87               & /                & 73               & 23               \\
\textbf{Education}                   & 69               & 2               & 9                 & 2                 & 2                 & /                 & 2                  & /                 & 42               & 2                & 123              & 2                \\
\textbf{Local Service}               & 9                & 4               & /                 & /                 & /                 & /                 & /                  & /                 & 3                & /                & 12               & /                \\
\textbf{Automotive Service}          & 18               & /               & 16                & /                 & 6                 & 2                 & 6                  & 2                 & 67               & /                & 21               & 5                \\
\textbf{Real Estate}                 & 128              & /               & 14                & /                 & 1                 & /                 & 1                  & /                 & 48               & /                & 17               & 1                \\
\textbf{Traffic}                     & 13               & /               & /                 & /                 & 2                 & /                 & 4                  & /                 & 3                & /                & 7                & /                \\
\textbf{Logistics}                   & 6                & /               & 1                 & /                 & 1                 & /                 & 1                  & /                 & 1                & /                & 10               & 4                \\
\textbf{Business}                    & 109              & /               & 25                & /                 & 13                & /                 & 13                 & /                 & 113              & /                & 82               & 4                \\
\textbf{Finance}                     & 2                & /               & 3                 & /                 & 2                 & /                 & 2                  & /                 & 1                & /                & 5                & /                \\
\textbf{Medical Service}             & 25               & /               & 14                & /                 & 2                 & /                 & 2                  & /                 & 30               & /                & 10               & 4                \\
\textbf{IT}                          & 2                & /               & 1                 & /                 & /                 & /                 & /                  & /                 & 1                & /                & /                & /                \\
\textbf{Social}                      & 1                & /               & /                 & /                 & 1                 & /                 & 1                  & /                 & 3                & /                & 7                & /                \\
\textbf{Government}                  & 28               & /               & 2                 & /                 & 5                 & /                 & 4                  & 3                 & 3                & /                & 10               & /                \\
\textbf{E-commerce}                  & 5                & /               & /                 & /                 & 2                 & /                 & 2                  & /                 & 6                & /                & 7                & /                \\
\textbf{Sports}                      & 6                & /               & /                 & /                 & /                 & /                 & /                  & /                 & 12               & /                & 3                & /                \\
\textbf{Dining}                      & 93               & /               & 60                & /                 & 57                & /                 & 57                 & /                 & 102              & /                & 99               & /                \\
\textbf{Entertainment}               & 4                & /               & /                 & /                 & 1                 & 1                 & 1                  & 1                 & /                & /                & 6                & /                \\
\textbf{Charity}                     & 1                & /               & /                 & /                 & /                 & /                 & /                  & /                 & /                & /                & 2                & /                \\
\textbf{Tourism}                     & 26               & /               & 8                 & /                 & 34                & /                 & 34                 & /                 & 36               & /                & 13               & /                \\
\textbf{News}                        & 11               & /               & 5                 & /                 & /                 & /                 & /                  & /                 & 15               & /                & 24               & /                \\ \hline
\textbf{Total}                       & \textbf{1,546}             & \textbf{20}              & \textbf{616}               & \textbf{4}                 & \textbf{858}               & \textbf{3}                 & \textbf{859}                & \textbf{9}                 & \textbf{2,815}            & \textbf{4}                & \textbf{1,155}            & \textbf{76}               \\
\hline
\end{tabular}
\end{table*}

\noindent\textbf{Web-View Domain Context Clustering.} 
Due to the same-origin policy restriction, cookie sharing typically occurs within the same domain, hence we employ the domain name as the basis for mini-program clustering.
\autoref{tab:classes} outlines the filtering process employed to identify mini-programs with web-views and cluster them based on their web-view domains. From the initial pool of 351,483 mini-programs, we first filtered for those containing the \texttt{<web-view>} component~(Filter 1). Next, we performed web-view domain clustering~(Filter 2), resulting in 45,448 groups of mini-programs with identifiable web-view domains. We then narrowed our focus to clusters containing more than 5 mini-programs~(Filter 3), leaving 3,109 groups. These clusters represent potential candidates for further analysis, as they may exhibit vulnerabilities associated with cross-component data sharing and communication through web-views under the same domain.

\noindent \textbf{Distribution of Web-View Domains.} The web-view domain clustering results are illustrated in \autoref{fig:overlap}. The green section represents the WeChat official domains, while the red section represents the third-party domains. The significantly larger size of the red section indicates that a considerable portion of the URLs in the dataset belong to third-party operators. Our analysis focused solely on grouping and investigating the domain names from the third-party operators. The rationale behind this approach is that the official domains associated with WeChat are trusted and unlikely to pose significant security risks. However, the third-party domains warrant closer examination, as they may potentially introduce vulnerabilities or be associated with unauthorized data sharing practices.

\noindent \textbf{Privileged Data Transmitted.}
The results of our analysis, as presented in \autoref{tab:privileged-data}, reveal widespread transmission of privileged data between mini-programs and their embedded web-views. Across all categories, we identified a total of 7,849 instances of sensitive data being sent from mini-programs to web-views, and 116 instances of data being received from web-views.
The sensitive data transmitted includes phone numbers, nicknames, location data~(latitude and longitude), avatar images, and user IDs. Among these, avatar images were the most frequently transmitted, with 2,815 instances, followed by user IDs (1,155 instances) and phone numbers (1,546 instances). Location data was also widely shared, with latitude and longitude transmitted in 858 and 859 instances, respectively. This extensive sharing of sensitive information not only exposes user privacy but also enables the possibility of tracking users across different mini-programs.
Certain categories of mini-programs exhibited particularly high levels of privileged data transmission. The \textit{Individual Seller} category was the most significant contributor, accounting for 772 instances of phone number transmission, 548 instances of location data sharing, and 1,783 instances of avatar image transmission. Other notable contributors include the \textit{Life Service} category, with 175 instances of location data sharing, and the \textit{Real Estate} category, which accounted for 128 instances of phone number transmission. 
While the majority of privileged data transmission occurred in a one-way manner, from mini-programs to web-views, the 116 instances of data being received from web-views highlight additional privacy risks. 

\noindent \textbf{Identification of Overlapping Web-View Domains.}  
Through our domain-based clustering and privileged data transmission analysis, we identified a total of 6,352 web-view domains that were involved in sending sensitive user data and 41 web-view domains that received such data. Among them, 7 domains exhibited both sending and receiving of sensitive data, indicating potential bi-directional data flows and higher risks of collusion attacks. These overlapping domains were further analyzed to assess their security and privacy implications.

\begin{table}[t]
\centering
\fontsize{8}{13}\selectfont
\begin{tabular}{lllr}
    \hline
    \textbf{Web-View Domain} & \textbf{Cookie} & \textbf{Data} & \textbf{Count}\\
    \hline
    \multirow{4}{*}{h5.youzan.com} & yz\_log\_uuid & user\_id & \multirow{4}{*}{5,952} \\
                                   & m & phone number &  \\
                                   & banner\_id & product id &  \\
                                   & yz\_log\_uid & user\_id &  \\
    \hline
    \multirow{2}{*}{watch-dog.huolala.cn} & device\_id & user\_id & \multirow{2}{*}{1,871} \\
                                          & user\_id & user\_id &  \\
    \hline
    m.jd.com & jd\_pin & username & 1,213 \\
    \hline
    \multirow{2}{*}{w.fooww.com} & nick\_name & username & \multirow{2}{*}{398} \\
                                  & headimgurl & avatar &  \\
    \hline
    \multirow{2}{*}{sf-express.com} & device\_id & user\_id & \multirow{2}{*}{279} \\
                                     & distinct\_id & user\_id &  \\
    \hline
    dinghuo123.com & distinct\_id & user\_id & 119 \\
    \hline
    m.ctrip.com & DUID & user\_id & 45 \\
    \hline
\end{tabular}
\caption{Collusion Attack via CMCS in the Wild.}
\label{tab:manual-analysis}
\end{table}

\noindent \textbf{Victim Analysis of Collusion Attacks.}
To understand the extent of the impact caused by cookie-sharing mechanisms, we analyzed the mini-programs affected by the 7 overlapping domains that exhibited both sending and receiving of sensitive data. Based on the data presented in \autoref{tab:manual-analysis}, a total of 9,877 mini-programs were identified as potential victims of cookie-sharing practices. These mini-programs were exposed to significant privacy risks due to the improper handling of sensitive user data.
Among the affected domains, \texttt{h5.youzan.com} had the greatest impact, involving 5,952 mini-programs that transmitted sensitive data such as phone numbers, product identifications, and user IDs through cookies, including \texttt{m}, \texttt{banner\_id}, and \texttt{yz\_log\_uid}. The domain \texttt{watch-dog.huolala.cn} affected 1,871 mini-programs, where device and user IDs were shared via cookies. Similarly, \texttt{m.jd.com} impacted 1,213 mini-programs through the sharing of usernames stored in the \texttt{jd\_pin} cookie. 
These findings highlight the urgent need for stricter security controls and enhanced data protection measures to mitigate these risks and prevent further harm.

\noindent \textbf{Findings.}  
Our measurement study reveals critical risks in the mini-program ecosystem. The integration of web-view components disrupts isolation mechanisms, enabling unauthorized data exchange and collusion attacks. We identified 7,965 instances of sensitive data transmission and uncovered 7 overlapping domains with bi-directional data flows, exposing 9,877 mini-programs to collusion attacks. These findings underscore the need for stronger isolation mechanisms, stricter scrutiny of web-view communication channels, and robust data protection policies to secure the ecosystem.
\section{Discussion}
\noindent \textbf{Generality of CMCS Vulnerability.} We also tested mini-programs other than WeChat. We found that the mini-programs of Alipay, Douyin, and Baidu also provide the same web-view component for embedding external web pages as the WeChat mini-program. We used the officially provided mini-program development tool to build a local test demo that embeds external web pages in a web-view, and used a Node.js-based web backend to set cookies. We found that these super applications behave similarly to WeChat. In the web-view of these mini-programs, there is no complete isolation between the view component and the embedded web page. Although we demonstrated and measured the collusion attack enabled by CMCS vulnerabilities, its potential impacts could extend far beyond this scenario. For example, an attacker could exploit CMCS by loading the same web-view as the victim's mini-program to impersonate the victim and log into their account without authorization. By inheriting the cookies stored in the shared web-view, the attacker could bypass authentication mechanisms and gain access to sensitive user data or perform malicious operations on behalf of the victim. These potential impacts underscore the urgent need for stricter isolation mechanisms in mini-program ecosystems.

\noindent \textbf{Risk Mitigation.}  
To address the risks associated with cookie sharing attacks and enhance security in the mini-program ecosystem, we propose several mitigation strategies:  
First and important,  isolate WebView instances per mini-program by default. Each WebView instance should be bound to the app ID of the mini-program that initiated it. This unique identifier would ensure that the WebView can only interact with the specific mini-program that launched it. By default, any attempt by a WebView to interact with other mini-programs should be blocked, thereby preventing unauthorized data sharing or cross-contamination.  
Additionally, for scenarios requiring cross-program interactions, we propose implementing an explicit user-consent mechanism. When a WebView initiated by a mini-program attempts to interact with another mini-program or access its data, the system should explicitly request the user's permission. This mechanism would clearly inform users of the intended data exchange and require their consent before proceeding, ensuring transparency and control.  
By implementing these measures, the mini-program ecosystem can achieve stronger isolation and greater transparency while maintaining flexibility for necessary interactions.

\noindent \textbf{Limitations.} While our study provides valuable insights into CMCS vulnerabilities, several limitations warrant consideration.
First, the collusion attack requires interaction with the attacker-controlled two mini-programs, which may reduce the feasibility of the attack in certain scenarios. However, we emphasize that this interaction does not need to occur simultaneously. The sender and receiver mini-programs can be used at different times, leveraging the persistent shared cookie storage managed by the global WebView process within the host application. 
Second, \tool is a large-scale static taint analysis framework that is path-insensitive but field-sensitive. It relies on predefined keywords to identify sensitive information, which means it does not provide strict guarantees of soundness and precision. Nevertheless, we believe that \tool is sufficiently effective for identifying CMCS risks in mini-program ecosystems. 
Third, \tool currently resolves dynamically constructed URLs through static data flow tracking across WXML and JavaScript files. This includes common patterns such as assignments via `getApp().globalData` and `setData`. However, if the URL is derived from runtime user input or dynamically fetched remote content, static analysis is unable to resolve it. While dynamic analysis could improve recall in such cases, we have not integrated it into \tool to maintain scalability for large-scale ecosystem analysis.

\section{Related Work}

\noindent \textbf{Security and Privacy of Mini-programs.} The security and privacy of mini-programs have garnered considerable attention in recent years~\cite{wang2023one,wang2023uncovering,zhang2024dark,zhang2024minicat,cai2025miniapp,yang2025minimal}, with various studies addressing different aspects of this emerging application paradigm. 
Notably, a significant body of work has focused on identifying and addressing security vulnerabilities within mini-programs. For instance, Wang et al.~\cite{wang2022webug} collected 83 real-world bugs and developed WeDetector, a tool that identifies WeBugs following three specific bug patterns. Another investigation~\cite{lu2020demystifying} explored issues such as system resource exposure, subwindow deception, and subapp lifecycle hijacking in the mini-program ecosystem, conducting evaluations on 11 popular platforms to highlight the prevalence of these security problems. Additionally, a novel privacy leak issue in mini-programs was studied~\cite{zhang2022identity}, which could potentially lead to private data theft by the mini-program platform, and an attack process exploiting this vulnerability was elucidated. Moreover, the Cross Miniapp Request Forgery (CMRF) vulnerability was identified~\cite{yang2022cmrf} in mini-program communication, shedding light on its root cause and designing the CMRFScanner tool for its detection. 
Beyond security vulnerabilities, a series of studies have also emphasized the importance of privacy in the mini-program ecosystem. For example, TaintMini~\cite{wang2023taintmini} introduced a framework for detecting flows of sensitive data within and across mini-programs using static taint analysis, while MiniTracker~\cite{li2023minitracker} constructed assignment flow graphs as a common representation across different host apps, revealing common privacy leakage patterns in a large-scale study of 150k mini-programs. Furthermore, several studies~\cite{zhang2023dont, baskaran2023measuring, meng2023wemint} have focused on taint analysis techniques to detect AppSecret leaks and some others~\cite{wang2023doasyousay, zhang2023spochecker, wang2024miniscope} focused on the consistency of data collection and usage in mini-programs, emphasizing the urgent
need for more precise privacy monitoring systems. 
Despite these significant contributions to understanding the privacy of mini-programs, there is still a need for further exploration and research. Different from the existing works, our study focused on both the user tracking and privacy risks of cross mini-program cookie sharing caused by the absence of web-view isolation in cross mini-program channels.

\noindent \textbf{Collusion Attacks.}
Collusion attacks, where multiple malicious applications collude to bypass security boundaries and gain unauthorized access to sensitive data, have been explored in the Android ecosystem. Several researches~\cite{bosu2017collusive,alhanahnah2019detecting,wang2023iafdroid} have investigated collusion attacks facilitated by different attack vectors. AppHolmes~\cite{xu2017appholmes} conducted the first in-depth study on app collusion in the mobile ecosystem, where one app covertly launches others in the background, leading to performance, efficiency, and security implications. Bosu et al.~\cite{bosu2017collusive} conducted a large-scale analysis of 110,150 Android apps to detect collusive and vulnerable apps based on inter-app ICC data flows, providing real-world evidence and insights into various types of ICC abuse for collusion attacks. Bhandari et al.~\cite{bhandari2017detecting} proposed a model-checking based approach for detecting collusion among Android apps by analyzing simultaneous inter-app communication and information leakage paths, improving upon prior techniques' detection capabilities.
It is important to note that while the existing research provides valuable insights into collusion attacks in various contexts, the unique characteristics and security models of the mini-program ecosystem necessitate tailored research and mitigation strategies. The introduction of shared web-view components and the interplay between mini-programs and web-view content pose distinct challenges that require specialized analysis and countermeasures.

% \noindent \textbf{Cookie Abusing and User Tracking.} 
% The abuse of cookies for user tracking and privacy violations has been extensively studied in the context of traditional web browsers and online advertising ecosystems~\cite{mayer2012third,papadogiannakis2021user,chen2021cookie,sivakorn2016cracked,munir2023cookiegraph}.
% Acar et al.~\cite{acar2014web} investigated the use of respawning and respawned cookies by popular websites, which can be exploited for user tracking even after cookie deletion. Englehardt et al.~\cite{englehardt2015cookies} conducted a large-scale measurement study of cookie practices on the web, highlighting the pervasiveness of tracking mechanisms and the potential for privacy violations. Their work shed light on the complex ecosystem of third-party trackers and the challenges in managing cookies effectively. Papadopoulos et al.~\cite{papadopoulos2019cookie} conducted an in-depth study of cookie synchronization (CSync) in the wild, revealing how this mechanism enables third-party trackers to merge user data and reconstruct browsing histories, bypassing the same-origin policy and exacerbating privacy concerns related to user tracking on the web. While prior research on cookie abusing explored mechanisms for merging user data across third-party trackers, our work focuses on the unique challenges posed by the mini-program ecosystem, where the integration of web-views and the interaction between mini-programs and web content create novel attack surfaces for cookie abuse and user tracking.

\section{Conclusion}
In this paper, we systematically investigated the security and privacy risks posed by CMCS vulnerabilities in the app-in-app ecosystem. Our analysis revealed that CMCS vulnerabilities exist in four major platforms, including WeChat, AliPay, TikTok, and Baidu, and can be exploited to perform collusion attacks.  
To address this issue, we proposed \tool{}, a static analysis tool that integrates web-view context modeling and cross-webview data flow analysis to detect CMCS vulnerabilities. Using \tool{}, we conducted the first large-scale measurement study of 351,483 real-world mini-programs, uncovering 45,448 clusters of mini-programs sharing web-view domains and 9,877 vulnerable mini-programs exposed to collusion attacks. These findings highlight the widespread impact of CMCS vulnerabilities and underscore the urgent need for improved isolation mechanisms and stricter data protection practices in mini-program ecosystems.  

% \section*{Artifact Availability}
% To promote transparency and reproducibility in our research, we have made the relevant artifacts publicly available at: \url{https://figshare.com/s/ffac7949a379e5f38401}.

\section*{Acknowledgement}
This work was supported in part by the National Natural Science Foundation of China (grants No.62572209, 62502168) and the Hubei Provincial Key Research and Development Program (grant No. 2025BAB057).
% \ifCLASSOPTIONcompsoc
%   % The Computer Society usually uses the plural form
%   \section*{Acknowledgments}
% \else
%   % regular IEEE prefers the singular form
% \fi

\bibliographystyle{IEEEtran}
\bibliography{reference}

\begin{thebibliography}{10}
\providecommand{\url}[1]{#1}
\csname url@samestyle\endcsname
\providecommand{\newblock}{\relax}
\providecommand{\bibinfo}[2]{#2}
\providecommand{\BIBentrySTDinterwordspacing}{\spaceskip=0pt\relax}
\providecommand{\BIBentryALTinterwordstretchfactor}{4}
\providecommand{\BIBentryALTinterwordspacing}{\spaceskip=\fontdimen2\font plus
\BIBentryALTinterwordstretchfactor\fontdimen3\font minus \fontdimen4\font\relax}
\providecommand{\BIBforeignlanguage}[2]{{%
\expandafter\ifx\csname l@#1\endcsname\relax
\typeout{** WARNING: IEEEtran.bst: No hyphenation pattern has been}%
\typeout{** loaded for the language `#1'. Using the pattern for}%
\typeout{** the default language instead.}%
\else
\language=\csname l@#1\endcsname
\fi
#2}}
\providecommand{\BIBdecl}{\relax}
\BIBdecl

\bibitem{lu2020demystifying}
H.~Lu, L.~Xing, Y.~Xiao, Y.~Zhang, X.~Liao, X.~Wang, and X.~Wang, ``Demystifying resource management risks in emerging mobile app-in-app ecosystems,'' in \emph{Proceedings of the 2020 ACM SIGSAC conference on computer and communications Security}, 2020, pp. 569--585.

\bibitem{yang2022cmrf}
Y.~Yang, Y.~Zhang, and Z.~Lin, ``Cross miniapp request forgery: Root causes, attacks, and vulnerability detection,'' in \emph{Proceedings of the 2022 ACM SIGSAC Conference on Computer and Communications Security}, 2022, pp. 3079--3092.

\bibitem{w3c}
W3C, ``Miniapp standardization white paper,'' 2022, \url{https://www.w3.org/TR/mini-app-white-paper}.

\bibitem{wang2023miniprogram}
\BIBentryALTinterwordspacing
S.~Wang, Y.~Zhao, K.~Wang, and H.~Wang, ``On the usage-scenario-based data minimization in mini programs,'' in \emph{Proceedings of the 2023 ACM Workshop on Secure and Trustworthy Superapps}, ser. SaTS '23.\hskip 1em plus 0.5em minus 0.4em\relax New York, NY, USA: Association for Computing Machinery, 2023, p. 29–32. [Online]. Available: \url{https://doi.org/10.1145/3605762.3624435}
\BIBentrySTDinterwordspacing

\bibitem{wang2024miniscope}
\BIBentryALTinterwordspacing
S.~Wang, Y.~Li, K.~Wang, Y.~Liu, H.~Li, Y.~Liu, and H.~Wang, ``Miniscope: Automated ui exploration and privacy inconsistency detection of miniapps via two-phase iterative hybrid analysis,'' \emph{ACM Trans. Softw. Eng. Methodol.}, Dec. 2024, just Accepted. [Online]. Available: \url{https://doi.org/10.1145/3709351}
\BIBentrySTDinterwordspacing

\bibitem{wang2024minichecker}
\BIBentryALTinterwordspacing
Y.~Wang, M.~Fan, H.~Zhou, H.~Wang, W.~Jin, J.~Li, W.~Chen, S.~Li, Y.~Zhang, D.~Han, and T.~Liu, ``Minichecker: Detecting data privacy risk of abusive permission request behavior in mini-programs,'' in \emph{Proceedings of the 39th IEEE/ACM International Conference on Automated Software Engineering}, ser. ASE '24.\hskip 1em plus 0.5em minus 0.4em\relax New York, NY, USA: Association for Computing Machinery, 2024, p. 1667–1679. [Online]. Available: \url{https://doi.org/10.1145/3691620.3695534}
\BIBentrySTDinterwordspacing

\bibitem{han2024policy}
\BIBentryALTinterwordspacing
Y.~Han, Z.~Xiao, Z.~Wang, and J.~Zhang, ``Privacy policy compliance in miniapps: An analytical study,'' in \emph{Proceedings of the ACM Workshop on Secure and Trustworthy Superapps}, ser. SaTS '24.\hskip 1em plus 0.5em minus 0.4em\relax New York, NY, USA: Association for Computing Machinery, 2024, p. 11–17. [Online]. Available: \url{https://doi.org/10.1145/3689941.3695777}
\BIBentrySTDinterwordspacing

\bibitem{deng2024counterfeit}
\BIBentryALTinterwordspacing
X.~Deng, M.~Zhang, X.~Dong, and X.~Hu, ``Detect counterfeit mini-apps: A case study on wechat,'' in \emph{Proceedings of the ACM Workshop on Secure and Trustworthy Superapps}, ser. SaTS '24.\hskip 1em plus 0.5em minus 0.4em\relax New York, NY, USA: Association for Computing Machinery, 2024, p. 1–10. [Online]. Available: \url{https://doi.org/10.1145/3689941.3695773}
\BIBentrySTDinterwordspacing

\bibitem{zhang2022identity}
L.~Zhang, Z.~Zhang, A.~Liu, Y.~Cao, X.~Zhang, Y.~Chen, Y.~Zhang, G.~Yang, and M.~Yang, ``Identity confusion in $\{$WebView-based$\}$ mobile app-in-app ecosystems,'' in \emph{31st USENIX Security Symposium (USENIX Security 22)}, 2022, pp. 1597--1613.

\bibitem{yang2023sok}
Y.~Yang, C.~Wang, Y.~Zhang, and Z.~Lin, ``{SoK: Decoding the Super App Enigma: The Security Mechanisms, Threats, and Trade-offs in OS-alike Apps},'' \emph{{arXiv preprint}}, 2023.

\bibitem{zhang2024minible}
\BIBentryALTinterwordspacing
Z.~Zhang, J.~Du, W.~Diao, and J.~Wu, ``Minible: Exploring insecure ble api usages in mini-programs,'' in \emph{Proceedings of the ACM Workshop on Secure and Trustworthy Superapps}, ser. SaTS '24.\hskip 1em plus 0.5em minus 0.4em\relax New York, NY, USA: Association for Computing Machinery, 2024, p. 18–22. [Online]. Available: \url{https://doi.org/10.1145/3689941.3695774}
\BIBentrySTDinterwordspacing

\bibitem{yang2023muid}
\BIBentryALTinterwordspacing
Z.~Yan, M.~Fan, Y.~Wang, J.~Shi, H.~Wang, and T.~Liu, ``Muid: Detecting sensitive user inputs in miniapp ecosystems,'' in \emph{Proceedings of the 2023 ACM Workshop on Secure and Trustworthy Superapps}, ser. SaTS '23.\hskip 1em plus 0.5em minus 0.4em\relax New York, NY, USA: Association for Computing Machinery, 2023, p. 17–21. [Online]. Available: \url{https://doi.org/10.1145/3605762.3624429}
\BIBentrySTDinterwordspacing

\bibitem{wang2024rootfree}
\BIBentryALTinterwordspacing
C.~Wang, Y.~Zhang, and Z.~Lin, ``Rootfree attacks: Exploiting mobile platform's super apps from desktop,'' in \emph{Proceedings of the 19th ACM Asia Conference on Computer and Communications Security}, ser. ASIA CCS '24.\hskip 1em plus 0.5em minus 0.4em\relax New York, NY, USA: Association for Computing Machinery, 2024, p. 830–842. [Online]. Available: \url{https://doi.org/10.1145/3634737.3645001}
\BIBentrySTDinterwordspacing

\bibitem{wang2023browser}
\BIBentryALTinterwordspacing
Y.~Wang, Y.~Yao, S.~Shi, W.~Chen, and L.~Huang, ``Towards a better super-app architecture from a browser security perspective,'' in \emph{Proceedings of the 2023 ACM Workshop on Secure and Trustworthy Superapps}, ser. SaTS '23.\hskip 1em plus 0.5em minus 0.4em\relax New York, NY, USA: Association for Computing Machinery, 2023, p. 23–28. [Online]. Available: \url{https://doi.org/10.1145/3605762.3624427}
\BIBentrySTDinterwordspacing

\bibitem{zhang2023trusted}
\BIBentryALTinterwordspacing
Z.~Zhang, Z.~Zhang, K.~Lian, G.~Yang, L.~Zhang, Y.~Zhang, and M.~Yang, ``Trusteddomain compromise attack in app-in-app ecosystems,'' in \emph{Proceedings of the 2023 ACM Workshop on Secure and Trustworthy Superapps}, ser. SaTS '23.\hskip 1em plus 0.5em minus 0.4em\relax New York, NY, USA: Association for Computing Machinery, 2023, p. 51–57. [Online]. Available: \url{https://doi.org/10.1145/3605762.3624430}
\BIBentrySTDinterwordspacing

\bibitem{han2023security}
\BIBentryALTinterwordspacing
Y.~Han, X.~Ji, Z.~Wang, and J.~Zhang, ``Systematic analysis of security and vulnerabilities in miniapps,'' in \emph{Proceedings of the 2023 ACM Workshop on Secure and Trustworthy Superapps}, ser. SaTS '23.\hskip 1em plus 0.5em minus 0.4em\relax New York, NY, USA: Association for Computing Machinery, 2023, p. 1–9. [Online]. Available: \url{https://doi.org/10.1145/3605762.3624432}
\BIBentrySTDinterwordspacing

\bibitem{android}
Android, ``Application sandbox,'' \url{https://source.android.com/docs/security/app-sandbox}, 2025, accessed: 2025-05-31.

\bibitem{sandbox}
WeChat, ``File system of wechat mini-programs,'' \url{https://developers.weixin.qq.com/minigame/en/dev/guide/base-ability/file-system.html}, 2025, accessed: 2025-05-31.

\bibitem{processiso}
DARKNAVY, ``Achieving persistent client-side attacks with a single wechat message,'' \url{https://www.darknavy.org/blog/achieving_persistent_client_side_attacks_with_a_single_wechat_message/}, 2025, accessed: 2025-05-31.

\bibitem{webview}
WeChat, ``Webview component in wechat mini-programs,'' \url{https://developers.weixin.qq.com/miniprogram/en/dev/component/web-view.html}, 2025, accessed: 2025-05-31.

\bibitem{wechat}
------, ``Mini program development platform,'' \url{https://developers.weixin.qq.com/miniprogram/dev/framework/}, 2025, accessed: 2025-05-31.

\bibitem{alipay}
AliPay, ``Mini program development platform,'' \url{https://miniprogram.alipay.com/}, 2025, accessed: 2025-05-31.

\bibitem{tiktok}
TikTok, ``Mini program development platform,'' \url{https://developer.open-douyin.com/docs/resource/zh-CN/mini-app/introduction/usage-guide}, 2025, accessed: 2025-05-31.

\bibitem{baidu}
Baidu, ``Mini program development platform,'' \url{https://smartprogram.baidu.com/docs/develop/tutorial/intro/}, 2025, accessed: 2025-05-31.

\bibitem{wang2023taintmini}
C.~Wang, R.~Ko, Y.~Zhang, Y.~Yang, and Z.~Lin, ``Taintmini: Detecting flow of sensitive data in mini-programs with static taint analysis,'' in \emph{Proceedings of the 45th International Conference on Software Engineering}, 2023.

\bibitem{zhang2021measurement}
Y.~Zhang, B.~Turkistani, A.~Y. Yang, C.~Zuo, and Z.~Lin, ``A measurement study of wechat mini-apps,'' \emph{Proceedings of the ACM on Measurement and Analysis of Computing Systems}, vol.~5, no.~2, pp. 1--25, 2021.

\bibitem{wang2023one}
C.~Wang, Y.~Zhang, and Z.~Lin, ``One size does not fit all: Uncovering and exploiting cross platform discrepant $\{$APIs$\}$ in $\{$WeChat$\}$,'' in \emph{32nd USENIX Security Symposium (USENIX Security 23)}, 2023, pp. 6629--6646.

\bibitem{wang2023uncovering}
------, ``Uncovering and exploiting hidden apis in mobile super apps,'' in \emph{Proceedings of the 2023 ACM SIGSAC Conference on Computer and Communications Security}, 2023, pp. 2471--2485.

\bibitem{zhang2024dark}
Z.~Zhang, L.~Zhang, G.~Yang, Y.~Chen, J.~Xu, and M.~Yang, ``The dark forest: Understanding security risks of cross-party delegated resources in mobile app-in-app ecosystems,'' \emph{IEEE Transactions on Information Forensics and Security}, 2024.

\bibitem{zhang2024minicat}
\BIBentryALTinterwordspacing
Z.~Zhang, Q.~Hou, L.~Ying, W.~Diao, Y.~Gu, R.~Li, S.~Guo, and H.~Duan, ``Minicat: Understanding and detecting cross-page request forgery vulnerabilities in mini-programs,'' in \emph{Proceedings of the 2024 on {ACM} {SIGSAC} Conference on Computer and Communications Security, {CCS} 2024, Salt Lake City, UT, USA, October 14-18, 2024}, B.~Luo, X.~Liao, J.~Xu, E.~Kirda, and D.~Lie, Eds.\hskip 1em plus 0.5em minus 0.4em\relax {ACM}, 2024, pp. 525--539. [Online]. Available: \url{https://doi.org/10.1145/3658644.3670294}
\BIBentrySTDinterwordspacing

\bibitem{cai2025miniapp}
\BIBentryALTinterwordspacing
Y.~Cai, Z.~Zhang, M.~Yao, J.~Liu, X.~Zhao, X.~Fu, R.~Li, Z.~Li, X.~Chen, Y.~Guo, and D.~Li, ``I can tell your secrets: Inferring privacy attributes from mini-app interaction history in super-apps,'' \emph{CoRR}, vol. abs/2503.10239, 2025. [Online]. Available: \url{https://doi.org/10.48550/arXiv.2503.10239}
\BIBentrySTDinterwordspacing

\bibitem{yang2025minimal}
\BIBentryALTinterwordspacing
Y.~Yang, Y.~Zhang, and Z.~Lin, ``Understanding miniapp malware: Identification, dissection, and characterization,'' in \emph{32nd Annual Network and Distributed System Security Symposium, {NDSS} 2025, San Diego, California, USA, February 24-28, 2025}.\hskip 1em plus 0.5em minus 0.4em\relax The Internet Society, 2025. [Online]. Available: \url{https://www.ndss-symposium.org/ndss-paper/understanding-miniapp-malware-identification-dissection-and-characterization/}
\BIBentrySTDinterwordspacing

\bibitem{wang2022webug}
T.~Wang, Q.~Xu, X.~Chang, W.~Dou, J.~Zhu, J.~Xie, Y.~Deng, J.~Yang, J.~Yang, J.~Wei \emph{et~al.}, ``Characterizing and detecting bugs in wechat mini-programs,'' in \emph{Proceedings of the 44th International Conference on Software Engineering}, 2022, pp. 363--375.

\bibitem{li2023minitracker}
W.~Li, B.~Yang, H.~Ye, L.~Xiang, Q.~Tao, X.~Wang, and C.~Zhou, ``Minitracker: Large-scale sensitive information tracking in mini apps,'' \emph{IEEE Transactions on Dependable and Secure Computing}, 2023.

\bibitem{zhang2023dont}
Y.~Zhang, Y.~Yang, and Z.~Lin, ``Don't leak your keys: Understanding, measuring, and exploiting the appsecret leaks in mini-programs,'' \emph{arXiv preprint arXiv:2306.08151}, 2023.

\bibitem{baskaran2023measuring}
S.~Baskaran, L.~Zhao, M.~Mannan, and A.~Youssef, ``Measuring the leakage and exploitability of authentication secrets in super-apps: The wechat case,'' \emph{arXiv preprint arXiv:2307.09317}, 2023.

\bibitem{meng2023wemint}
S.~Meng, L.~Wang, S.~Wang, K.~Wang, X.~Xiao, G.~Bai, and H.~Wang, ``Wemint: Tainting sensitive data leaks in wechat mini-programs,'' in \emph{2023 38th IEEE/ACM International Conference on Automated Software Engineering (ASE)}.\hskip 1em plus 0.5em minus 0.4em\relax IEEE, 2023, pp. 1403--1415.

\bibitem{wang2023doasyousay}
Y.~Wang, M.~Fan, J.~Liu, J.~Tao, W.~Jin, H.~Wang, Q.~Xiong, and T.~Liu, ``Do as you say: Consistency detection of data practice in program code and privacy policy in mini-app,'' \emph{IEEE Transactions on Software Engineering}, vol.~50, no.~12, pp. 3225--3248, 2024.

\bibitem{zhang2023spochecker}
X.~Zhang, Y.~Wang, X.~Zhang, Z.~Huang, L.~Zhang, and M.~Yang, ``Understanding privacy over-collection in wechat sub-app ecosystem,'' \emph{arXiv preprint arXiv:2306.08391}, 2023.

\bibitem{bosu2017collusive}
A.~Bosu, F.~Liu, D.~Yao, and G.~Wang, ``Collusive data leak and more: Large-scale threat analysis of inter-app communications,'' in \emph{Proceedings of the 2017 ACM on Asia Conference on Computer and Communications Security}, 2017, pp. 71--85.

\bibitem{alhanahnah2019detecting}
M.~Alhanahnah, Q.~Yan, H.~Bagheri, H.~Zhou, Y.~Tsutano, W.~Srisa-An, and X.~Luo, ``Detecting vulnerable android inter-app communication in dynamically loaded code,'' in \emph{IEEE INFOCOM 2019-IEEE Conference on Computer Communications}.\hskip 1em plus 0.5em minus 0.4em\relax IEEE, 2019, pp. 550--558.

\bibitem{wang2023iafdroid}
B.~Wang, C.~Yang, and J.~Ma, ``Iafdroid: Demystifying collusion attacks in android ecosystem via precise inter-app analysis,'' \emph{IEEE Transactions on Information Forensics and Security}, 2023.

\bibitem{xu2017appholmes}
M.~Xu, Y.~Ma, X.~Liu, F.~X. Lin, and Y.~Liu, ``Appholmes: Detecting and characterizing app collusion among third-party android markets,'' in \emph{Proceedings of the 26th international conference on World Wide Web}, 2017, pp. 143--152.

\bibitem{bhandari2017detecting}
S.~Bhandari, F.~Herbreteau, V.~Laxmi, A.~Zemmari, P.~S. Roop, and M.~S. Gaur, ``Detecting inter-app information leakage paths,'' in \emph{Proceedings of the 2017 ACM on Asia Conference on Computer and Communications Security}, 2017, pp. 908--910.

\end{thebibliography}

\end{document}